\documentclass[twocolumn,superscriptaddress,groupedaddress,footinbib]{revtex4}
\usepackage{graphicx}
\usepackage{subfigure}
\usepackage{amsmath}
\usepackage{comment}
\usepackage{color}

\begin{document}

\title{Degenerate Mobilities in Phase Field Models are Insufficient to Capture Surface Diffusion}

\author{Alpha A Lee}
\affiliation{Mathematical Institute, University of Oxford, Oxford OX2 6GG, United Kingdom}

\author{Andreas M{\"u}nch}
\affiliation{Mathematical Institute, University of Oxford, Oxford OX2 6GG, United Kingdom}

\author{Endre S\"uli}
\affiliation{Mathematical Institute, University of Oxford, Oxford OX2 6GG, United Kingdom}

\begin{abstract}
Phase field models frequently provide insight to phase transitions, and are robust numerical tools to solve free boundary problems corresponding to the motion of interfaces. A body of prior literature suggests that interface motion via surface diffusion is the long-time, sharp interface limit of microscopic phase field models such as the Cahn-Hilliard equation with a degenerate mobility function.  Contrary to this conventional wisdom, we show that the long-time behaviour of degenerate Cahn-Hilliard equation with a polynomial free energy undergoes coarsening, reflecting the presence of bulk diffusion, rather than pure surface diffusion. This reveals an important limitation of phase field models that are frequently used to model surface diffusion.  
\end{abstract}

\makeatother
\maketitle

A key problem in modelling phase transitions in materials lies in linking macroscopic interfacial motion and mesoscopic dynamics. A common approach are phase field models, which replace a sharp interface with order parameters that are continuous across the interface. Phase field models can be constructed from systematic coarse graining of the microscopic Hamiltonian, and are often written as a gradient minimisation of certain microscopic free energy functional \cite{anderson1998diffuse,chen2002phase,provatas2011phase}. As such, they provide the crucial link between microscopic interactions and the kinetics of phase separation and pattern-formation. 

Mesocopic phase field descriptions have been widely used in the literature as their numerical approximation is less complicated than the approximation of macroscopic descriptions based on sharp interfaces. By using a continuous order parameter field, phase field approaches are versatile enough to capture topological changes, and replace the numerically challenging task of interface tracking with integration of a time-dependent partial differential equation. Therefore, the phase field formalism is increasingly used as a numerical approximation for a wider class of free boundary problems than what the free energy describes microscopically \cite{elder2001sharp,emmerich2008advances,steinbach2009phase}. 

A particularly noteworthy class of free boundary problem is when the velocity of the interface $v_n$ is proportional to the surface Laplacian of the mean curvature $\kappa$ and
\begin{equation}
v_n = \mathcal{M} \Delta_s \kappa, 
\label{vel_SD}
\end{equation}
where $\mathcal{M}$ is the mobility. Equation (\ref{vel_SD}) is known as the surface diffusion flow, and has been used as a model for many complex processes such as electromigration in metals \citep{mahadevan1999phase}, heteroepitaxial growth \citep{ratz2006surface} and more recently solid-solid dewetting \citep{jiang2012phase}. 

The Cahn-Hilliard equation with degenerate mobility is the commonly used phase field model to approximate surface diffusion (e.g. \cite{kitahara1978kinetic, mahadevan1999phase,bhate2000diffuse,yeon2006phase,ratz2006surface,wise2007solving,torabi2009new,jiang2012phase}), where the order parameter $u$ is conserved and satisfies (in dimensionless units)
\begin{subequations}
\begin{align}
       u_t &= -\nabla \cdot \mathbf{j}, \; \;\; \mathbf{j} = -\epsilon M(u) \nabla \mu, \label{CHE_1} \\ 
        \epsilon \mu &= -\epsilon^2 \nabla^2 u + f'(u),   \label{eq_mu} \\    
        f(u) &= \frac{1}{4}(1-u^2)^2,  \quad M(u) = 1-u^2, 
\end{align}
\label{deg_CHE}%
\end{subequations} 
where $M(u)$ is the mobility function; $\mathbf{j}$ is the flux; $\mu$ is the chemical potential; $\epsilon$ is the interfacial tension which determines the width of the interface, and $f(u)$ is the bulk free energy. Throughout the paper, we will assume the no-flux condition $\mathbf{n}\cdot \mathbf{j} =0 $, and the variational Neumann condition $\mathbf{n} \cdot \nabla u = 0$ on the boundaries of the solution domain. The two pure phases are denoted by $u = \pm 1$, and the interface is located at the contour $u=0$. 

The precise form of the mobility is usually chosen on thermodynamic grounds \cite{giacomin1996exact,nauman2001nonlinear}. For the lattice-gas entropy $s(u) = u \log u + (1-u)\log(1-u)$, the Einstein relation stipulates that the mobility is related to the entropy function $s(u)$ via $M(u) = (\partial^2 s/\partial u^2)^{-1} = 1-u^2$. This motivates the choice of mobility in (\ref{deg_CHE}). 

We are interested in the long-time behaviour when the initial mixture
has separated into regions where $u$ is either close to $1$ or to $-1$,
except for regions of width $O(\epsilon)$
close to the interface over which $u$ transitions between these two regions.
Heuristically, the width of this interface layer decreases as $\epsilon \rightarrow 0 $. One may think that if the mobility function is degenerate and vanishes at the pure phases, the flux normal to the interface is suppressed and therefore only surface diffusion via mass flux tangential to the interface can occur. 
However, this heuristic argument neglects the fact that the gradient of the interface diverges as $\epsilon \rightarrow 0$. Therefore, whether the degenerate mobility function is sufficient to suppress the normal mass flux at leading order is unclear \cite{BrayE95,gugenberger2008comparison}. 

\begin{figure}
\centerline{\includegraphics[scale=0.3]{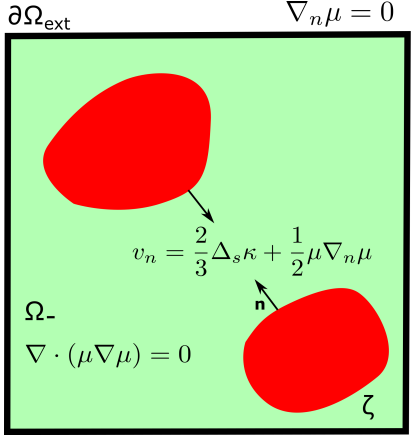}}
\caption{Illustration of the asymptotic structure of the degenerate
mobility case.  The domain $\Omega_{\mathrm{ext}}$ is split into interfaces
$ \zeta$ where $u=0$ which enclose the solid domains $\Omega_+$ where $u>0$
(coloured red), and the region $\Omega_-$ with vapor (colored green) where
$u>0$.  The normal $\mathbf{n}$ to the interfaces points out of $\Omega_+$.}
\label{asym_structure}
\end{figure}

The key result of this paper is the presence of a nonlinear bulk diffusion term at leading order for the interface velocity, and the correct sharp interface limit that describes the
quasistationary evolution of the interface $\zeta$, located at $u=0$, is given by 
\begin{subequations}
\label{new_porous_medium}
\begin{align}
\nabla \cdot (\mu \nabla \mu) &= 0, \quad \text{in } \Omega_-,  
\label{new_porous_mediuma} \\ 
\mu &= \frac{2}{3} \kappa, \quad \text{on }  \zeta,  
\label{new_porous_mediumb} \\ 
v_n &= \frac{2}{3}  \Delta_s  \kappa + \frac{1}{4} \mu \nabla_n \mu \quad \text{on }  \zeta, 
\label{new_porous_mediumc} \\
\nabla_n \mu &= 0 , \quad \text{on }  \partial \Omega_{\mathrm{ext}},
\label{new_porous_mediumd}
\end{align}
\end{subequations}
where $v_n$ is the normal velocity, and the definitions of $\Omega_-$ and $\zeta$ are given in Figure \ref{asym_structure}. The possibility of bulk diffusion was noted in earlier works \cite{BrayE95,CahnT94} and analysed recently by Dai and Du \cite{dai2014coarsening}. However, in \cite{dai2014coarsening} the analysis was done on the (unphysical) solution branch with $|u|>1$ in some region. Our analysis below considers the physical branch of solution where $|u|<1$ everywhere, and derives the limiting model as $\epsilon \rightarrow 0$. It can be shown using rigorous mathematical analysis that the physical branch with $|u|<1$ everywhere exists for all parameter values \cite{elliott1996cahn}. 


To begin our analysis, we drop the time derivative from
(\ref{deg_CHE}).  Rewriting the Laplacian in polar coordinates, and resolving the boundary layer near the interface
by noting that the interface is $O(\epsilon)$ thick, we obtain the
leading order solution
\begin{equation}
u_0(r) = \tanh \left( \frac{r-r_0}{\epsilon} \right),  \quad \mu=\mu\big|_\zeta.
\label{stat}
\end{equation} 
Here $r$ is the radial coordinate with respect to the centre of the
osculating circle to the interface, $r_0$ is the position of the
interface in these coordinates, and $\mu\big|_\zeta$ is a constant. Equation
(\ref{stat}) reveals how the interface is being represented as a
continuous, albeit thin, order parameter profile of width $\epsilon$
around $r=r_0$. A key physical intuition is that the relaxation of the local order parameter profile to (\ref{stat}) is rapid (as it is driven by local rearrangement of particles), and the late stage dynamics is quasi-static and determined by the movement of the interface, \emph{i.e.} change in $r_0$. 

To put this intuition onto firmer footing, we consider the chemical potential $\mu$ away from the interface. There, the system is almost a pure phase (say $u = -1 + \epsilon \tilde{u}$, $\mu = \epsilon \tilde{\mu}$, with $\tilde{u},\tilde{\mu} = O(1)$), and the spatial variation of $u$ is negligible so that the Laplacian can be neglected. Expanding Equation (\ref{eq_mu}) in powers $\epsilon$, the leading order term is given by 
\begin{equation} 
\tilde{\mu} =  f''(-1)\tilde{u},
\label{mu_assym}
\end{equation}
and substituting (\ref{mu_assym}) into (\ref{CHE_1}), we obtain 
\begin{equation} 
\tilde{u}_t = M'(-1) \nabla\cdot( \tilde{u} \nabla \tilde{\mu} ) = \frac{M'(-1)}{f''(-1)} \nabla\cdot( \tilde{\mu} \nabla \tilde{\mu} ). 
\label{outer}
\end{equation}
Further, we use the aforementioned quasi-static approximation, and assume that $\tilde{u}_t$ is small. Thus, to leading order, 
\begin{equation}
\nabla\cdot( \tilde{\mu} \nabla \tilde{\mu} ) = 0. 
\end{equation}  

The chemical potential at the interface can be obtained by multiplying (\ref{eq_mu}) by $\partial_r u$, and integrating both sides. Assuming that the azimuthal variations in $u$ is asymptotically smaller than the radial variation, we have 
\begin{equation}
\int_{r_0 - \eta}^{r_0 +  \eta} \epsilon \mu \partial_r u \mathrm{d}r = \int_{r_0 -  \eta}^{r_0 +  \eta}  \;  \left[ - \frac{\epsilon^2}{r} \partial_r (r \partial_r u) + f'(u) \right] \partial_r u \; \mathrm{d}r. 
\label{eq_for_mu}
\end{equation}
Close to the interface, $u$ can be approximated by $u(r) \approx u_0(r)$. Substituting (\ref{stat}) into (\ref{eq_for_mu}), and assuming $\epsilon \ll \eta \ll 1$, we obtain 
\begin{equation}
\mu\big|_\zeta = \frac{2}{3} \kappa, 
\end{equation}
where $\kappa = 1/r_0$ is the curvature of the interface. 

In the quasi-static approximation, we neglect time dependence except for the
slow motion of the interface.
To obtain the interface velocity, we focus on the boundary layer region, and move into a Lagrangian frame by making the transformation $u_t \mapsto u_t - (\mathbf{v} \cdot \nabla) u$. Now, noting that normal velocity is much larger than the lateral velocity, we have $(\mathbf{v} \cdot \nabla) u \approx \epsilon^2 v_n  \partial_{r} u$, where $\epsilon^2 v_n$ is the normal velocity,
scaled in anticipation of the order of the right hand side which will
turn out to be $O(\epsilon^2)$. Equation (\ref{CHE_1}) becomes 
\begin{equation}
- \epsilon^2 v_n  \partial_{r} u = - \partial_r (\mathbf{j} \cdot \mathbf{n}) - 
\partial_s (\mathbf{j} \cdot \mathbf{t}),
\end{equation}
where $s$ denotes the coordinate tangent to the interface, and $ \mathbf{n}$ and $ \mathbf{t}$ 
are the (outward pointing) unit normal and the unit tangent vector, respectively.  
Close to the interface, the tangential flux is given by $ \mathbf{j} \cdot \mathbf{t} =
-\epsilon M(u) \partial_s \mu = -\epsilon M(u) \partial_s \mu|_\zeta $.
Thus, again substituting (\ref{stat}) for $u$, and integrating over the interface, we arrive at
\begin{equation}
2 \epsilon^2 v_n  = - \left[\mathbf{j} \cdot \mathbf{n}\right]_{r_0-\eta}^{r_0 + \eta} 
+ \frac43\epsilon^2 \partial_{ss} \kappa. 
\end{equation} 
Now, for  $\epsilon\ll\eta\ll1$, we obtain
\[
\mathbf{j} \cdot \mathbf{n} \Big|_{r_0 + \eta} = 
-\epsilon M(u) \partial_r \mu=-\epsilon^2\frac{M'(-1)}{f''(-1)}  \mu \partial_r  \mu + O(\epsilon^3)
\]
and $\mathbf{j} \cdot \mathbf{n} \big|_{r_0 - \eta} =0$ as the system reaches the pure phase $u=1$ deep in $\Omega_+$.
Therefore, all in all, we obtain \eqref{new_porous_mediumc}, 
where we have identified $\partial_{ss}$ with the surface Laplacian $\Delta_{s}$. Equation (\ref{new_porous_mediumc}) shows that the interface velocity in the sharp interface model has two contributions: one from surface diffusion, Equation (\ref{vel_SD}), which is local to the interface, and another contribution from nonlinear bulk diffusion  
which satisfies a porous-medium equation \eqref{new_porous_mediuma}. Unlike pure surface diffusion,
the mass flux arising from bulk diffusion couples
disjoint solid domains with each other.  This results in coarsening where larger solid domains
grow at the expense of smaller ones, which cannot happen for pure surface diffusion. 
Moreover, for non-circular interfaces, the contribution from surface and bulk diffusion
enter the interface evolution to the same order, i.e.\ the effect of the latter 
does not become negligible compared to the former even when letting $\epsilon\to0$.


To test this prediction of our analysis, we consider the relaxation of an
azimuthal perturbation to a radially symmetric stationary state with radius $r_0$ 
and hence curvature
$\kappa = 1/r_0$.  For azimuthal perturbations proportional to $\cos m\theta$, the pure
solid diffusion model (\ref{vel_SD}) predicts an exponential decay rate
\begin{equation}
\sigma = - \frac{2}{3} \kappa^4 m^2 (m^2-1). 
\label{pure_solid_diffusion_decay_rate}
\end{equation} 
In contrast, the decay rate in the porous medium model (\ref{new_porous_medium}) is given by
\begin{equation}
\sigma = - \kappa^4 \left[ \frac{2}{3} m^2 (m^2-1) + \frac{1}{9} m (m^2-1) \tanh(m \log \kappa) \right].
\label{solid_diffusion_one_side_decay_rate}
\end{equation}
Table \ref{diffuse_sharp_deg1} shows the decay rate numerically obtained
by solving the phase field model, Equation (\ref{deg_CHE}).  It shows
how the decay rate of the azimuthal perturbation to the axisymmetric base
state tends to the linearised sharp
interface model \emph{with} with the contribution from nonlinear bulk diffusion,
rather than to the one for pure surface diffusion.

\begin{table}
\centering
\begin{tabular}{| l | l | l | l | l | l | l |}
\hline 
 $\epsilon$ & 0.01 &  0.005  &  0.002 & 0.001  & \bf Eq (\ref{solid_diffusion_one_side_decay_rate}) & \bf Eq (\ref{pure_solid_diffusion_decay_rate}) \\ \hline
 $\sigma$ & $-$133.2  & $-$133.8  & $-$136.3 & $-$137.0 & \bf $-$137.4 
& \bf $-$128 \\ 
 \hline
\end{tabular}
\caption{Relaxation rates for $m=2$ obtained from the linearised phase field model are shown for different values of $\epsilon$ in the first four columns, and compared to the eigenvalues obtained for linearised sharp interface models for pure surface diffusion (\ref{pure_solid_diffusion_decay_rate}) and the porous medium type model (\ref{solid_diffusion_one_side_decay_rate}) in the next-to-last and the last column, respectively. }
\label{diffuse_sharp_deg1}
\end{table} 

Our analysis only applies to free energy functions $f$ for which $f'(\pm 1)=0$
and $f''(\pm 1)>0$ at the minima $u=\pm 1$.
Indeed,  according to the asymptotic
analysis in
\cite{cahn1996cahn}, the double obstacle free energy 
\begin{equation*}
f_{\mathrm{do}}(u) = \begin{cases} 1-u^2 &\text{if } |u| < 1, \\ 
\infty & \text{if } |u| \geq  1  \end{cases} 
\end{equation*}
gives rise to pure surface diffusion flow for $M (u) = 1-u^2$ in the sharp
interface limit ($\epsilon\to0$), and also
if the logarithmic free energy $f(u) = - u^2  + \epsilon^\alpha (u \log u + (1-u) \log(1-u))$
with $\alpha>0$ is used instead. 

In conclusion, our analysis establishes that phase field models 
for pure surface surface diffusion cannot be realised
using the Cahn-Hilliard equation with the degenerate mobility  
and Ginzburg-Landau free energy as in \eqref{deg_CHE}, as
was repeatedly assumed in the literature  \cite{kitahara1978kinetic, wise2007solving, torabi2009new, jiang2012phase}. A nonlinear bulk diffusion term
appears to leading order of the sharp interface limit, hence affecting
the coarsening behaviour on the same time scale as surface diffusion.
In particular, it allows, on this time scale, 
disjoint interfaces to coarsen and cannot be suppressed by reducing $\epsilon$.

We note that the derivation presented in this work could be made mathematically robust via matched asymptotic analysis. Such an approach was applied to analyse the Cahn-Hilliard equation with constant mobility \cite{pego1989front,alikakos1994convergence}.  Extending the method of matched asymptotics to model (\ref{deg_CHE}) with a degenerate mobility will be the subject of a subsequent publication. The heuristic approach presented here, however, reveals clearly the salient physics involved in the sharp interface limit. 

\bibliography{ref_cahn_hilliard}

\begin{thebibliography}{24}
\expandafter\ifx\csname natexlab\endcsname\relax\def\natexlab#1{#1}\fi
\expandafter\ifx\csname bibnamefont\endcsname\relax
  \def\bibnamefont#1{#1}\fi
\expandafter\ifx\csname bibfnamefont\endcsname\relax
  \def\bibfnamefont#1{#1}\fi
\expandafter\ifx\csname citenamefont\endcsname\relax
  \def\citenamefont#1{#1}\fi
\expandafter\ifx\csname url\endcsname\relax
  \def\url#1{\texttt{#1}}\fi
\expandafter\ifx\csname urlprefix\endcsname\relax\def\urlprefix{URL }\fi
\providecommand{\bibinfo}[2]{#2}
\providecommand{\eprint}[2][]{\url{#2}}

\bibitem[{\citenamefont{Anderson et~al.}(1998)\citenamefont{Anderson, McFadden,
  and Wheeler}}]{anderson1998diffuse}
\bibinfo{author}{\bibfnamefont{D.}~\bibnamefont{Anderson}},
  \bibinfo{author}{\bibfnamefont{G.~B.} \bibnamefont{McFadden}},
  \bibnamefont{and} \bibinfo{author}{\bibfnamefont{A.}~\bibnamefont{Wheeler}},
  \bibinfo{journal}{Annual review of fluid mechanics}
  \textbf{\bibinfo{volume}{30}}, \bibinfo{pages}{139} (\bibinfo{year}{1998}).

\bibitem[{\citenamefont{Chen}(2002)}]{chen2002phase}
\bibinfo{author}{\bibfnamefont{L.-Q.} \bibnamefont{Chen}},
  \bibinfo{journal}{Annual review of materials research}
  \textbf{\bibinfo{volume}{32}}, \bibinfo{pages}{113} (\bibinfo{year}{2002}).

\bibitem[{\citenamefont{Provatas and Elder}(2011)}]{provatas2011phase}
\bibinfo{author}{\bibfnamefont{N.}~\bibnamefont{Provatas}} \bibnamefont{and}
  \bibinfo{author}{\bibfnamefont{K.}~\bibnamefont{Elder}},
  \emph{\bibinfo{title}{Phase-field methods in materials science and
  engineering}} (\bibinfo{publisher}{John Wiley \& Sons},
  \bibinfo{year}{2011}).

\bibitem[{\citenamefont{Elder et~al.}(2001)\citenamefont{Elder, Grant,
  Provatas, and Kosterlitz}}]{elder2001sharp}
\bibinfo{author}{\bibfnamefont{K.}~\bibnamefont{Elder}},
  \bibinfo{author}{\bibfnamefont{M.}~\bibnamefont{Grant}},
  \bibinfo{author}{\bibfnamefont{N.}~\bibnamefont{Provatas}}, \bibnamefont{and}
  \bibinfo{author}{\bibfnamefont{J.}~\bibnamefont{Kosterlitz}},
  \bibinfo{journal}{Physical Review E} \textbf{\bibinfo{volume}{64}},
  \bibinfo{pages}{021604} (\bibinfo{year}{2001}).

\bibitem[{\citenamefont{Emmerich}(2008)}]{emmerich2008advances}
\bibinfo{author}{\bibfnamefont{H.}~\bibnamefont{Emmerich}},
  \bibinfo{journal}{Advances in Physics} \textbf{\bibinfo{volume}{57}},
  \bibinfo{pages}{1} (\bibinfo{year}{2008}).

\bibitem[{\citenamefont{Steinbach}(2009)}]{steinbach2009phase}
\bibinfo{author}{\bibfnamefont{I.}~\bibnamefont{Steinbach}},
  \bibinfo{journal}{Modelling and Simulation in Materials Science and
  Engineering} \textbf{\bibinfo{volume}{17}}, \bibinfo{pages}{073001}
  (\bibinfo{year}{2009}).

\bibitem[{\citenamefont{Mahadevan and Bradley}(1999)}]{mahadevan1999phase}
\bibinfo{author}{\bibfnamefont{M.}~\bibnamefont{Mahadevan}} \bibnamefont{and}
  \bibinfo{author}{\bibfnamefont{R.~M.} \bibnamefont{Bradley}},
  \bibinfo{journal}{Physica D: Nonlinear Phenomena}
  \textbf{\bibinfo{volume}{126}}, \bibinfo{pages}{201} (\bibinfo{year}{1999}).

\bibitem[{\citenamefont{R{\"a}tz et~al.}(2006)\citenamefont{R{\"a}tz, Ribalta,
  and Voigt}}]{ratz2006surface}
\bibinfo{author}{\bibfnamefont{A.}~\bibnamefont{R{\"a}tz}},
  \bibinfo{author}{\bibfnamefont{A.}~\bibnamefont{Ribalta}}, \bibnamefont{and}
  \bibinfo{author}{\bibfnamefont{A.}~\bibnamefont{Voigt}},
  \bibinfo{journal}{Journal of Computational Physics}
  \textbf{\bibinfo{volume}{214}}, \bibinfo{pages}{187} (\bibinfo{year}{2006}).

\bibitem[{\citenamefont{Jiang et~al.}(2012)\citenamefont{Jiang, Bao, Thompson,
  and Srolovitz}}]{jiang2012phase}
\bibinfo{author}{\bibfnamefont{W.}~\bibnamefont{Jiang}},
  \bibinfo{author}{\bibfnamefont{W.}~\bibnamefont{Bao}},
  \bibinfo{author}{\bibfnamefont{C.~V.} \bibnamefont{Thompson}},
  \bibnamefont{and} \bibinfo{author}{\bibfnamefont{D.~J.}
  \bibnamefont{Srolovitz}}, \bibinfo{journal}{Acta Materialia}
  \textbf{\bibinfo{volume}{60}}, \bibinfo{pages}{5578} (\bibinfo{year}{2012}).

\bibitem[{\citenamefont{Kitahara and Imada}(1978)}]{kitahara1978kinetic}
\bibinfo{author}{\bibfnamefont{K.}~\bibnamefont{Kitahara}} \bibnamefont{and}
  \bibinfo{author}{\bibfnamefont{M.}~\bibnamefont{Imada}},
  \bibinfo{journal}{Progress of Theoretical Physics Supplement}
  \textbf{\bibinfo{volume}{64}}, \bibinfo{pages}{65} (\bibinfo{year}{1978}).

\bibitem[{\citenamefont{Bhate et~al.}(2000)\citenamefont{Bhate, Kumar, and
  Bower}}]{bhate2000diffuse}
\bibinfo{author}{\bibfnamefont{D.~N.} \bibnamefont{Bhate}},
  \bibinfo{author}{\bibfnamefont{A.}~\bibnamefont{Kumar}}, \bibnamefont{and}
  \bibinfo{author}{\bibfnamefont{A.~F.} \bibnamefont{Bower}},
  \bibinfo{journal}{Journal of Applied Physics} \textbf{\bibinfo{volume}{87}},
  \bibinfo{pages}{1712} (\bibinfo{year}{2000}).

\bibitem[{\citenamefont{Yeon et~al.}(2006)\citenamefont{Yeon, Cha, and
  Grant}}]{yeon2006phase}
\bibinfo{author}{\bibfnamefont{D.-H.} \bibnamefont{Yeon}},
  \bibinfo{author}{\bibfnamefont{P.-R.} \bibnamefont{Cha}}, \bibnamefont{and}
  \bibinfo{author}{\bibfnamefont{M.}~\bibnamefont{Grant}},
  \bibinfo{journal}{Acta Materialia} \textbf{\bibinfo{volume}{54}},
  \bibinfo{pages}{1623} (\bibinfo{year}{2006}).

\bibitem[{\citenamefont{Wise et~al.}(2007)\citenamefont{Wise, Kim, and
  Lowengrub}}]{wise2007solving}
\bibinfo{author}{\bibfnamefont{S.}~\bibnamefont{Wise}},
  \bibinfo{author}{\bibfnamefont{J.}~\bibnamefont{Kim}}, \bibnamefont{and}
  \bibinfo{author}{\bibfnamefont{J.}~\bibnamefont{Lowengrub}},
  \bibinfo{journal}{Journal of Computational Physics}
  \textbf{\bibinfo{volume}{226}}, \bibinfo{pages}{414} (\bibinfo{year}{2007}).

\bibitem[{\citenamefont{Torabi et~al.}(2009)\citenamefont{Torabi, Lowengrub,
  Voigt, and Wise}}]{torabi2009new}
\bibinfo{author}{\bibfnamefont{S.}~\bibnamefont{Torabi}},
  \bibinfo{author}{\bibfnamefont{J.}~\bibnamefont{Lowengrub}},
  \bibinfo{author}{\bibfnamefont{A.}~\bibnamefont{Voigt}}, \bibnamefont{and}
  \bibinfo{author}{\bibfnamefont{S.}~\bibnamefont{Wise}},
  \bibinfo{journal}{Proceedings of the Royal Society A: Mathematical, Physical
  and Engineering Science} \textbf{\bibinfo{volume}{465}},
  \bibinfo{pages}{1337} (\bibinfo{year}{2009}).

\bibitem[{\citenamefont{Giacomin and Lebowitz}(1996)}]{giacomin1996exact}
\bibinfo{author}{\bibfnamefont{G.}~\bibnamefont{Giacomin}} \bibnamefont{and}
  \bibinfo{author}{\bibfnamefont{J.~L.} \bibnamefont{Lebowitz}},
  \bibinfo{journal}{Physical Review Letters} \textbf{\bibinfo{volume}{76}},
  \bibinfo{pages}{1094} (\bibinfo{year}{1996}).

\bibitem[{\citenamefont{Nauman and He}(2001)}]{nauman2001nonlinear}
\bibinfo{author}{\bibfnamefont{E.~B.} \bibnamefont{Nauman}} \bibnamefont{and}
  \bibinfo{author}{\bibfnamefont{D.~Q.} \bibnamefont{He}},
  \bibinfo{journal}{Chemical Engineering Science}
  \textbf{\bibinfo{volume}{56}}, \bibinfo{pages}{1999} (\bibinfo{year}{2001}).

\bibitem[{\citenamefont{Bray and Emmott}(1995)}]{BrayE95}
\bibinfo{author}{\bibfnamefont{A.~J.} \bibnamefont{Bray}} \bibnamefont{and}
  \bibinfo{author}{\bibfnamefont{C.~L.} \bibnamefont{Emmott}},
  \bibinfo{journal}{Physical Review B} \textbf{\bibinfo{volume}{52}},
  \bibinfo{pages}{R685} (\bibinfo{year}{1995}).

\bibitem[{\citenamefont{Gugenberger et~al.}(2008)\citenamefont{Gugenberger,
  Spatschek, and Kassner}}]{gugenberger2008comparison}
\bibinfo{author}{\bibfnamefont{C.}~\bibnamefont{Gugenberger}},
  \bibinfo{author}{\bibfnamefont{R.}~\bibnamefont{Spatschek}},
  \bibnamefont{and} \bibinfo{author}{\bibfnamefont{K.}~\bibnamefont{Kassner}},
  \bibinfo{journal}{Physical Review E} \textbf{\bibinfo{volume}{78}},
  \bibinfo{pages}{016703} (\bibinfo{year}{2008}).

\bibitem[{\citenamefont{Cahn and Taylor}(1994)}]{CahnT94}
\bibinfo{author}{\bibfnamefont{J.}~\bibnamefont{Cahn}} \bibnamefont{and}
  \bibinfo{author}{\bibfnamefont{J.}~\bibnamefont{Taylor}},
  \bibinfo{journal}{Acta Metallurgica et Materialia}
  \textbf{\bibinfo{volume}{42}}, \bibinfo{pages}{1045} (\bibinfo{year}{1994}),
  ISSN \bibinfo{issn}{0956-7151}.

\bibitem[{\citenamefont{Dai and Du}(2014)}]{dai2014coarsening}
\bibinfo{author}{\bibfnamefont{S.}~\bibnamefont{Dai}} \bibnamefont{and}
  \bibinfo{author}{\bibfnamefont{Q.}~\bibnamefont{Du}},
  \bibinfo{journal}{Multiscale Modeling \& Simulation}
  \textbf{\bibinfo{volume}{12}}, \bibinfo{pages}{1870} (\bibinfo{year}{2014}).

\bibitem[{\citenamefont{Elliott and Garcke}(1996)}]{elliott1996cahn}
\bibinfo{author}{\bibfnamefont{C.~M.} \bibnamefont{Elliott}} \bibnamefont{and}
  \bibinfo{author}{\bibfnamefont{H.}~\bibnamefont{Garcke}},
  \bibinfo{journal}{SIAM Journal on Mathematical Analysis}
  \textbf{\bibinfo{volume}{27}}, \bibinfo{pages}{404} (\bibinfo{year}{1996}).

\bibitem[{\citenamefont{Cahn et~al.}(1996)\citenamefont{Cahn, Elliott, and
  Novick-Cohen}}]{cahn1996cahn}
\bibinfo{author}{\bibfnamefont{J.~W.} \bibnamefont{Cahn}},
  \bibinfo{author}{\bibfnamefont{C.~M.} \bibnamefont{Elliott}},
  \bibnamefont{and}
  \bibinfo{author}{\bibfnamefont{A.}~\bibnamefont{Novick-Cohen}},
  \bibinfo{journal}{European Journal of Applied Mathematics}
  \textbf{\bibinfo{volume}{7}}, \bibinfo{pages}{287} (\bibinfo{year}{1996}).

\bibitem[{\citenamefont{Pego}(1989)}]{pego1989front}
\bibinfo{author}{\bibfnamefont{R.~L.} \bibnamefont{Pego}},
  \bibinfo{journal}{Proceedings of the Royal Society of London. A. Mathematical
  and Physical Sciences} \textbf{\bibinfo{volume}{422}}, \bibinfo{pages}{261}
  (\bibinfo{year}{1989}).

\bibitem[{\citenamefont{Alikakos et~al.}(1994)\citenamefont{Alikakos, Bates,
  and Chen}}]{alikakos1994convergence}
\bibinfo{author}{\bibfnamefont{N.~D.} \bibnamefont{Alikakos}},
  \bibinfo{author}{\bibfnamefont{P.~W.} \bibnamefont{Bates}}, \bibnamefont{and}
  \bibinfo{author}{\bibfnamefont{X.}~\bibnamefont{Chen}},
  \bibinfo{journal}{Archive for Rational Mechanics and Analysis}
  \textbf{\bibinfo{volume}{128}}, \bibinfo{pages}{165} (\bibinfo{year}{1994}).

\end{thebibliography}

\end{document}